\documentclass{article}[12pt]

\begin{document}

\newcommand{\pref}[1]{(\ref{#1})}

\newcommand{\bra}[1]{\mbox{$\langle{#1}|$}}
\newcommand{\ket}[1]{\mbox{$|{#1}\rangle$}}
\newcommand{\bk}[2]{\mbox{$\langle{#1}|{#2}\rangle$}}
\newcommand{\braket}[3]{\mbox{$\langle{#1}|\it{#2}|{#3}\rangle$}}

\newcommand{\colt}[3]
{\mbox{$\left(\begin{array}{c}
{#1} \\ {#2} \\ {#3} \end{array}\right) $}}

\newcommand{\rowt}[3]
{\mbox{$\left(\begin{array}{ccc} #1 & #2 & #3 \end{array}\right)$}}

\newcommand{\matt}[9]
{\mbox{$\left(\begin{array}{ccc}
#1 & #2 & #3 \\ #4 & #5 & #6 \\ #7 & #8 & #9 \end{array}\right)$}}

\newcommand{\col}[2]
{\mbox{$\left(\begin{array}{c}
#1 \\ #2 \end{array}\right)$}}

\newcommand{\row}[2]
{\mbox{$\left(\begin{array}{cc}
#1 & #2 \end{array}\right)$}}

\newcommand{\mat}[4]
{\mbox{$\left(\begin{array}{cc}
#1 & #2 \\ #3 & #4 \end{array}\right)$}}

\newcommand{\txxbox}[1]
{\mbox{$\left[\begin{array}{c}
#1 \end{array}\right]$}}

\newcommand{\be}{\begin{equation}}

\newcommand{\ee}{\end{equation}}

\newcommand{\vc}[1]{\mbox{$\vec{{\bf #1}}$}}

\newcommand{\del}{\partial}

\newcommand{\pd}[1]{\frac{\del}{\del #1}}

\newcommand{\pdd}[2]{\frac{\del^2}{\del #1 \del #2}}

\newcommand{\Dd}[1]{\frac{d}{d #1}}

\input{epsf}

\begin{flushright}
EFI-97-50

hep-th/9710242
\end{flushright}

\begin{center}
\vspace{2cm}

{\Large {\bf On the Bound States of Matrix Strings}}

\vspace{3cm}

Vatche Sahakian

{\em Enrico Fermi Institute, Department of Physics

University of Chicago

5640 S. Ellis Ave., Chicago, IL 60637, USA }

\verb+isaak@maxwell.uchicago.edu+

\vspace{3cm}

\begin{minipage}{12cm}
\begin{center} {\large {\bf Abstract}} \end{center}
\vspace{1cm}

We investigate excitations in Matrix Theory on $T^2$ corresponding
to bound states of strings. We demonstrate the Dirichlet aspect of
R-R charged vacua through a non-trivial connection between the $U(1)$ and
$SU(n)$ sectors of the matrix SYM.
\end{minipage}

\end{center}
\vspace{3cm}
October 1997

\vspace{6pt}
\hrule

\newpage
\setcounter{page}{1}
\section{Introduction}

During the past few years, a web of perturbative and non-perturbative
connections has been cast between the different string theories of a decade
ago\cite{SchLect}; the common theme is that a limiting regime
of a given theory is described by another.
Probing the non-perturbative regime of the IIA string theory, with
$11$ dimensional supergravity as a description of its low energy
dynamics, M-theory\cite{WDyn} stands at a crucial junction in the web of string
dualities, with the need of a proper description
of its degrees of freedom. The BFSS conjecture\cite{Conj}
proposes that the full dynamics of M theory in the infinite momentum 
frame is captured by the matrix SYM describing IIA D0 branes. 
Given the already established connections
between M and other theories in various settings,
this proposal faces a multitude of consistency tests.
And many it has already, and remarkably, passed\cite{State}. Of relevance
to the current discussion, it is
expected, on the basis of proposed dualities, to account for the bound states
of IIB strings\cite{Wbound}\cite{SL2Z} in this Matrix Theory.
It was argued in~\cite{Motl}\cite{MST} that, 
Matrix Theory on $T^1$, a $1+1$ SYM,
metamorphoses, in the strong YM coupling regime, into the CFT
of the IIA string in the infinite momentum frame; this entailed
an elegant mechanism of ``screwing'' string-like excitations
into matrices. 
The first quantized 
IIB fundamental string was identified along a similar mechanism
in Matrix Theory on $T^2$ in~\cite{SeiStrings}; it was proposed that
the $SL(2,Z)$ S-duality of the IIB theory can then be used to realize
the spectrum of the IIB bound strings. The purpose of this paper is to
investigate this latter point in greater detail. 

We will study the energetics of
excitations in Matrix Theory on the torus. Storing NS-NS and R-R
charges in the $U(1)$ sector, we will propose to
wind the dual gauge scalar of the
theory on a cycle determined by the charge content of this vacuum.
The energetics of excitations in the $SU(n)$
will be seen to be affected by the $U(1)$ configuration
in a non-trivial pattern; quanta parallel to the $U(1)$ electric
field will become light if the latter corresponds to an R-R charged
configuration;
these light excitations will be indicative
of massive excitations of fundamental strings attached to the background,
a signature of the Dirichlet aspect of 
these vacua\cite{RRPolch}\cite{TASIPolch}. The heavy excitations
in these scenarios will be perpendicular to the electric field
and will be identified with the non-BPS excitations
of IIB bound strings. 

Throughout, we will be sloppy in
handling factors of $2\pi$; the reader may set $2\pi \rightarrow 1$ to
avoid grief.
 
\section{Matrix Theory on $T^d$}

Consider Matrix Theory as M theory in the infinite momentum frame described
by the quantum mechanics Lagrangian\cite{Conj}
\be
L=\frac{1}{2R_{11}}tr \dot{X}_i^2 -\frac{R_{11}}{2} 
\frac{1}{l_P^6}tr[X_i,X_j]^2+fermions \ ,
\ee
where the 9 $X$'s are hermitian $N\times N$ matrices and 
the quantum commutator is given by
\be
[X^i_{ab},\dot{X}^j_{cd}]=i R_{11} \delta_{ij} \delta_{ac} \delta_{bd}\ .
\ee

The dimensional reduction of this theory on a transverse torus $T^d$ can be
obtained by realizing a subset of the $U(N)$ gauge symmetry as the 
compactification of $d$ scalars, yielding a $d+1$ SYM\cite{Conj}\cite{Taylor}. 
After rescaling fields and coordinates, the form we will use is
\be
\mathcal{L}=\frac{1}{2}\left(\frac{l_P^2}{R_{11}}\right)^{3-d} 
tr \left \{
-\frac{F_{\mu \nu}^2}{(R_\mu R_\nu)^2} + 
\frac{(D_\mu X_i)^2}{R_\mu^2} - G^2
[X_i,X_j]^2 + fermions
\right \}\ ,
\ee
with
\be
F_{\mu\nu}\equiv R_\nu \partial_\mu A_\nu -R_\mu \partial_\nu A_\mu -i
R_\mu R_\nu G [A_\mu,A_\nu]
\ee
\be
D_\mu\equiv \partial_\mu -iR_\mu G [A_\mu,.]\ ,
\ee
where $\mu,\nu=0,\ldots ,d$, and $i,j=d+1,\ldots ,9$; note that repeated
indices in some of these equations are not summed over.
The dimensionless coupling $G$ is
\be
G^2\equiv \frac{l_P^d}{\prod_\mu R_\mu} \ .
\ee
The quantum commutator for the gauge or scalar fields becomes
\be
[X^i_{ab}(\sigma),\dot{X}^j_{cd}(\sigma')]=
\left( \frac{l_P^2}{R_{11}}\right)^{d-3} 
\prod_\mu \frac{\delta(\sigma^\mu-{\sigma'}^\mu)}{R_\mu} \delta_{ij}
\delta_{ac} \delta_{bd} \ .
\ee
This YM is on the torus parametrized by $\sigma^\mu$ for $\mu=1,\ldots ,d$,
with periods
\be
\Sigma^\mu=\frac{l_P^3}{R_{11} R_\mu^2} \label{periods}\ ,
\ee
and the measure of the action integral is
\be
d\tau \prod_\mu d\sigma^\mu R_\mu\ .
\ee
Some of the equations have time $\tau\equiv 
\sigma^0$ rescaled by an $R^0\sim 1$; no significance beyond notational convenience 
will be accorded to this scale. The fermions are ignored throughout;
supersymmetry can be invoked to reintroduce
their contributions to the final results. 

We will also need some of the charges appearing in the SUSY algebra\cite{MST}; 
in our notation, these are
\be
\oint tr F_{0 \mu}= w q_\mu \frac{R_{11} R_0}{G R_\mu^2} \label{charge1}
\ee
\be
\oint tr F_{\mu \nu}=\frac{w_{\mu \nu}}{G} \label{charge2}\ ,
\ee
where $w$, $q_\mu$ and $w_{\mu \nu}$ are integers by momentum and/or
Dirac quantization;
$w q_\mu$ is momentum in the appropriate direction, or Kaluza charge
equivalently, while $w_{\mu \nu}$ counts the winding 
of the $11D$ membrane on corresponding 2-cycles. We have also factored out
a common integer $w$ in the momenta; in the case $d=2$, this implies
that $q_1$ and $q_2$ are relatively prime.

We will consider this system for $d=2$, i.e. SYM $2+1$ as M-theory on 
a transverse torus. Taking $R_{11}\rightarrow \infty$ as part of
the infinite momentum frame prescription, 
the three length scales of this theory are $R_1$, $R_2$ and
$l_P$. We focus on the strong coupling regime of the YM
\be
G^2 \gg 1 \Rightarrow R_1 R_2 \ll l_P^2\ .
\ee
Energy considerations then justify dropping excitations in non-commuting
matrices; we then go to the diagonal gauge  where we have 
the residual gauge freedom corresponding to 
the permutations of matrix eigenvalues\cite{Motl}\cite{MST}. 
Furthermore, we can now
dualize the field strength~\cite{SeiStrings} 
into the derivative of a scalar $*F=d\phi$; 
we will do this
in the coordinates $R_\mu \sigma_\mu$; rescaling the scalar as well
for convenience, we define the duality map
\be
\dot{\phi}=\frac{G R_0}{R_1 R_2} F_{12}
\ee
\be
\partial_1 \phi=\frac{G R_1}{R_0 R_2} F_{02}
\ee
\be
\partial_2 \phi =-\frac{G R_2}{R_0 R_1} F_{01}\ .
\ee

In the regime we are considering, the Hamiltonian then takes the form
\be
H=\frac{1}{2 R_{11}}\int d^2\sigma 
tr \left(
R_1^2 (\partial_2 \phi)^2 + R_2^2 (\partial_1 \phi)^2 +
R_1^2 R_2^2 \dot{\phi}^2
+ R_1^2 (\partial_2 X^i)^2 + R_2^2 (\partial_1 X^i)^2 +
R_1^2 R_2^2 \dot{X^i}^2
\right)\ , \label{mainH}
\ee
where $i=3,\ldots ,9$, and 
the $X$'s have been rescaled $X\rightarrow X/G$; the quantum
commutator for either the $\phi$ or the $X$'s is
\be
[\phi_{aa}(\sigma),\dot{\phi}_{bb}(\sigma')]=i R_{11}
\frac{\delta^2(\sigma-{\sigma'})}{R_1^2 R_2^2} \delta_{ab}\ . \label{qcomm}
\ee

\section{Exciting matrices}

We want to look for excitations in the Hamiltonian of equation~\pref{mainH}
corresponding to the bound states of fundamental and 
Dirichlet strings, predicted to be in the spectrum through the duality
between M-theory on $T^2$ and the IIB theory on the circle\cite{SchDual}.
We focus on a block diagonal matrix configuration of size $n$ siting in
the larger $N\times N$ matrices; the size of this block is expected
to be identified with the longitudinal momentum of excitations it will carry,
and is to be taken to infinity.

We excite zero modes in the $U(1)$ gauge fields according to equations~\pref{charge1}
and \pref{charge2};
we then have an
electric fields along the $(q_1,q_2)$ cycle and a 2-form magnetic field
on the torus.
It easy to show that excitations in the
$U(1)$ apart from these zero modes
contribute mass $\propto n\rightarrow \infty$
\footnote{This can be seen as follows; the non-zero modes in the $U(1)$
will have period $\Sigma$; two derivatives in the energy and the trace
of the unit matrix contribute collectively a factor of $n$ to the Hamiltonian;
with the longitudinal momentum given by $p_+=n/R_{11}$, this corresponds
to a mass $\propto n$, which is out of the spectrum in the infinite
momentum frame.}.
Throwing away such non-zero modes in the $U(1)$, we find that the scalar
is given by
\be
\phi^{U(1)}=\frac{w_{12}}{n} \frac{R_{11}^2 R_1 R_2 R_0}{l_P^6} \tau
+\frac{w}{n} q_2 \frac{R_{11}^2 R_1}{l_P^3 R_2} \sigma^1
-\frac{w}{n} q_1 \frac{R_{11}^2 R_2}{l_P^3 R_1} \sigma^2\ . \label{eq1}
\ee
Along with zero modes in the $X$'s, this gives the $U(1)$ energy
\be
H^{U(1)}=\frac{R_{11}}{2n} \left( w^2 \left( \frac{q_1^2}{R_1^2} +
\frac{q_2^2}{R_2^2}\right)+
\left( \frac{w_{12} R_1 R_2}{l_P^3} \right)^2 +p_i^2 \right)\ .
\ee

The charges stored in the $U(1)$ will eventually be interpreted as R-R
or NS-NS charges in the IIB theory. Given the connection between R-R charges
and the 
Dirichlet aspect of objects carrying such charges, we expect a non-trivial
mechanism through which the $U(1)$ sector of the YM affects the
nature of the excitations in the $SU(n)$ sector; this despite the fact that,
superficially, the two sectors are decoupled. We will demonstrate this
connection by looking at the theory on a transformed torus.

We apply the transformation to new coordinates 
$(\tilde{\sigma}^1, \tilde{\sigma}^2)$
that aligns the one-form electric
field along $\tilde{\sigma}^1$; this
transformation then must take the form
\be
\col{\tilde{\sigma}^1}{\tilde{\sigma}^2}=\mat{q_1}{q_2}{c}{-d} 
\col{\sigma^1}{\sigma_2}\ , \label{transf}
\ee
where we restrict $c$ and $d$ by orthogonality and area preservation
\be
c=-\frac{q_2 R_1^2}{\Delta}
\ee
\be
d=-\frac{q_1 R_2^2}{\Delta}
\ee
\be
\Delta\equiv q_1^2 R_2^2 + q_2^2 R_1^2\ .
\ee
The $SU(n)$ part of the Hamiltonian of equation~\pref{mainH} becomes
\be
H^{SU(n)}=\frac{R_1 R_2}{2 R_{11}} \int d^2 \tilde{\sigma}
tr \left( 
\frac{\Delta}{R_1 R_2} (\tilde{\partial}_1 \phi)^2 + 
\frac{R_1 R_2}{\Delta} (\tilde{\partial}_2 \phi)^2
+ R_1 R_2 \dot{\phi}^2 + \mbox{similar X terms} \right)\ .
\ee
In the new variables, the scalar field looks like
\be
\phi^{U(1)}=\frac{w_{12}}{n}\left( \frac{R_{11}}{l_P^3}\right)^2 R_1 R_2 \tau
-\frac{w}{n} \frac{R_{11}^2 \Delta}{l_P^3 R_1 R_2} \tilde{\sigma}^2
\label{phiu1}\ .
\ee
Given that this zero mode structure depends on the 
coordinate $\tilde{\sigma}^2$, 
the natural candidate for string-like excitations away 
from this configuration are modes of the same $\sigma$ dependence, i.e.
perpendicular to the direction of the electric field
\be
\phi=\phi^{U(1)}\left(\tilde{\sigma}^2\right)+
\phi^{SU(n)}\left(\tilde{\sigma}^2\right)\ . \label{addeq}
\ee
To find the periods of the new torus, we will again make use of the
$U(1)$ sector. The scalar being compact 
with radius $R$, we wind $\tilde{\sigma}^2$ on R using equation~\pref{phiu1}
\footnote{We note that the transformation aligns the new axis along
integer valued vectors on the torus, and therefore yields to new periodic
coordinates; this point was brought to my attention by M. Li. We
also note that the momenta
stored in the $U(1)$ sector are quantized on the original torus, as seen
in equation~\pref{eq1}; in this respect, the transformation at hand is used as 
a tool to make manifest the energetics of the excitations.}
\be
\frac{R_{11}^2 \Delta}{l_P^3 R_1 R_2} \tilde{\Sigma}^2=R\ . \label{u1sun}
\ee
From the quantum commutator~\pref{qcomm}, we also have
\be
R=\frac{R_{11}}{R_1 R_2}\ ,
\ee
and area preservation implies
\be
\tilde{\Sigma}_1 \tilde{\Sigma}_2=\Sigma_1 \Sigma_2\ .
\ee
Putting things together, we find
\be
\tilde{\Sigma}_1=\frac{l_P^3 \Delta}{R_{11} R_1^2 R_2^2}
\ee
\be
\tilde{\Sigma}_2=\frac{l_P^3}{R_{11} \Delta}\ .
\ee
To analyze the energetics of the $SU(n)$ Hamiltonian, we rescale the
coordinates $(\tilde{\sigma}^1, \tilde{\sigma}^2)=
(\tilde{\Sigma}^1 {{\tilde{\sigma'}}}^1, \tilde{\Sigma}^2 {{\tilde{\sigma'}}}^2)$
to new variables of period 1 (or $2\pi$).
The $SU(n)$ Hamiltonian then takes the form
\begin{eqnarray}
H^{SU(n)}=\frac{1}{2} \int d^2 {{\tilde{\sigma}}'}
tr\left( R_1 R_2 \left(\frac{R_1 R_2}{\Delta}\right) \left( \frac{R_{11}}{l_P^3}
\right)^2 (\tilde{\partial'}_1 \phi)^2 \right. \nonumber \\
\left. + R_1 R_2 \left(\frac{\Delta}{R_1 R_2}\right) \left( \frac{R_{11}}{l_P^3}
\right)^2 (\tilde{\partial'}_2 \phi)^2 +
\dot{\phi}^2 + \cdots \right) \ , \label{maineq}
\end{eqnarray}
where we have rescaled the fields again so as to get the canonical form
of the commutator for the $\phi$, i.e.
schematically $[\phi,\dot{\phi}]=i\delta^2$.

\begin{figure}
\epsfxsize=11cm \centerline{\leavevmode \epsfbox{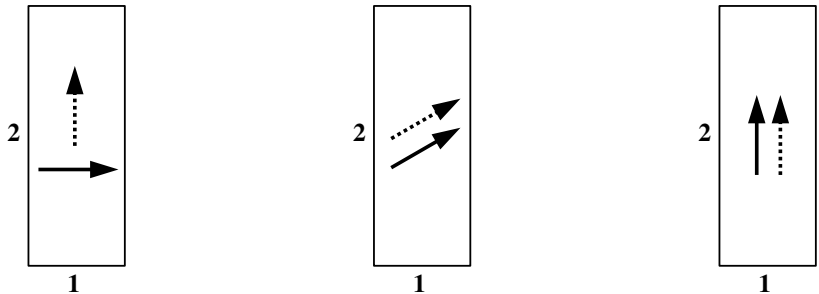}}
\caption{The YM torus in the regime $R_2\ll R_1$ 
(see equation~\pref{periods} for this identification); the solid arrow
is the direction of the $U(1)$ electric field in three different
scenarios, from left to right corresponding to $q_2=0$,
both charges nonzero, and $q_1=0$; the dotted arrow is the direction of
light excitations in each case. }
\label{fig1}
\end{figure}

Let us first consider the case where $q_1\neq 0$ and $q_2\neq 0$.
We then note that the following strict inequality holds
\be
\frac{\Delta}{R_1 R_2}>1 \Rightarrow \tilde{\Sigma}^1 > \tilde{\Sigma}^2
\label{ineq}
\ee
and can be pushed to the extreme by either $R_2\ll R_1$ or
$R_1\ll R_2$. The electric field is parallel to $\tilde{\sigma}^1$;
looking at the relative relations of the $\sigma$ derivatives 
in equation~\pref{maineq},
we notice that the string-like $SU(n)$ excitations
parallel to the $U(1)$ electric field are energetically favored over
the more natural candidates perpendicular to it, in
{\em both} regimes $R_2\ll R_1$ and $R_1\ll R_2$; 
we have the non-trivial statement that, in the skewed torus limit,
the identification of the direction of the
light $SU(n)$ excitations depends on the direction of the
$U(1)$ electric field, independent of our choice of the small radius;
the content of the $U(1)$ vacuum is affecting the nature of the excitations
in the $SU(n)$ due to the existence of a scalar in the theory winding
on a circle as in equation~\pref{u1sun}.
Extending this analysis to the zero charge cases as well, 
the different possible scenarios are illustrated in 
Figure~\ref{fig1}. We note the pattern that, whenever the electric field,
indicated by the solid arrow, projects a component along the long side
of the torus (i.e. the direction of the small radius), the light excitations
flip parallel to it, whereas the natural string-like excitations throughout are
expected to be perpendicular to the electric field, 
as seen from equation~\pref{addeq}.
We will demonstrate that this structure
translates into the link between
R-R charge and the Dirichlet aspect of R-R charged vacua.

For the purpose of
substantiating the meaning of both the light and heavy excitations
in the $SU(n)$ sector,
we would like to consider exciting each of the two 
cycles independently and writing down the corresponding energy spectra.
For the $U(1)$ sector, the zero modes are determined and we have the boundary
condition
\be
\phi^{U(1)}(\tilde{\sigma}^2+\tilde{\Sigma}^2)=
\phi^{U(1)}(\tilde{\sigma}^2) + R\frac{w}{n}
\ee
which was used in equation~\pref{u1sun} already.
Next, look at excitations in the $SU(n)$ with Motl-like twisted boundary
conditions~\cite{Motl} along either direction 
$\tilde{\sigma}^1$ or $\tilde{\sigma}^2$.
It can be shown that the most general physically relevant
boundary condition in the $SU(n)$ sector is
\be
\phi^{SU(n)}(\sigma+\Sigma)=g \phi^{SU(n)}(\sigma) g^{-1} \ , \label{bc}
\ee
where $\sigma$ is either $\tilde{\sigma}^1$ or $\tilde{\sigma}^2$
and $g$ is the shift operator in $Z_n$
\footnote{Elements of $S_N$ split into cycles in $Z_n$ 
such that, by construction,
we have focused on a block diagonal matrix mapping into 
one of these cycles;
see \cite{MST} for details on this construction.}; after
rescaling the time appropriately, we then easily find the form
\footnote{It is mildly interesting to note that
the structure of sewing the strings into these matrices can be shown to
forbid zero modes in the $SU(n)$. Generally, the residual gauge freedom in the
diagonal gauge consists of matrices of the form $g d(\sigma)$, where
$g \in S_N$ and is independent of $\sigma$ by continuity, while $d(\sigma)$
is a diagonal matrix contributing only through the derivative term in the
gauge transformation. The most general boundary
condition on the scalar is then
\be
\phi^{SU(n)}=g\phi^{SU(n)}g^{-1} +D \ ,
\ee
where D is a diagonal traceless matrix independent of $\sigma$ and
g is the shift operator in $Z_n$ as argued in the previous footnote. Writing
$\phi^{SU(n)}=M_1 + M_2 \sigma + M_3 \tau +\phi^{SU(n)}_{nonzero}$, we
can easily show that $M_2=M_3=0$, while $M_1$ can be gauged away. We are then
left with the most general boundary condition of equation~\pref{bc}
on the non-zero modes.}
\be
\phi^{SU(n)}= i \frac{\alpha_m}{m} \omega^m \exp{-\frac{2\pi m i}{\Sigma n}
(\tau-\sigma)} + i \frac{\tilde{\alpha}
_m}{m} \omega^{-m} \exp{-\frac{2\pi m i}{\Sigma n}
(\tau+\sigma)}
\ee
with $\omega=diag(e^{2\pi i/j})$. Similar boundary conditions and solutions
follow for the other $X$ polarizations.  The $SU(n)$ energies for excitations
along $\tilde{\sigma}^1$ and $\tilde{\sigma}^2$ are easily read off the
Hamiltonian~\pref{maineq}
\be
H^{SU(n)}_{(1)}=\frac{R_{11}}{2n} 4\pi \frac{\sqrt{R_1 R_2}}{l_P^3}
\left( q_1^2 \frac{R_2}{R_1} + q_2^2 \frac{R_1}{R_2} \right)^{-1/2}\hat{N} 
\ee
\be
H^{SU(n)}_{(2)}=\frac{R_{11}}{2n} 4\pi \frac{\sqrt{R_1 R_2}}{l_P^3}
\left( q_1^2 \frac{R_2}{R_1} + q_2^2 \frac{R_1}{R_2} \right)^{1/2}\hat{N} \ .
\ee
The $\hat{N}$ operator is the usual right and left number operator
which may now be thought to include contributions from fermionic excitations.

Consider the regime
\be
R_2 \ll R_1 \Rightarrow
R_2 \ll l_P\ .
\ee
From the point of view of M-theory and its known connections with IIA and
IIB string theories, excitations in this regime
are expected to be described by a
perturbative IIA theory with dilaton vev $e^{\phi_A}=g_A$ and 
string length scale $l_S$
given by
\be
R_2=g_A l_S
\ee
\be
l_P^3=l_S^3 g_A
\ee
or by a perturbative IIB theory with dilaton vev $e^{\phi_B}=g_B$
\be
g_B=R_2/R_1\ .
\ee
The IIB fundamental string tension is then given in terms of 
M theory parameters by
\be
T_F=\frac{R_2}{l_P^3} \ ,
\ee
while the IIB Dirichlet string tension is
\be
T_D=\frac{R_1}{l_P^3} \gg T_F\ .
\ee
Furthermore, $q_1$ gets interpreted as NS-NS
charge, while $q_2$ counts R-R charge. Looking at
Figure~\ref{fig1}, we note that
the configurations with light excitations parallel 
to the electric field correspond to scenarios carrying non-zero R-R charge. 

Consider first the case of $q_1=1$,$q_2=0$. 
The $SU(n)$ sector is excited energetically favorably in the
$\sigma^2$ direction, perpendicular to the electric field, 
as shown in Figure~\ref{fig1}. Given that
$H^{U(1)} \sim \mathcal{O}(T_F)$, the skewed torus limit singles out
the light string-like excitations to $\mathcal{O}(T_F)$ in the $SU(n)$
as well. The excitations off the $U(1)$ vacuum 
are 1+1 dimensional and the spectrum of the Hamiltonian $H=H^{U(1)} +H^{SU(n)}$ 
is that of the IIB fundamental string\cite{SeiStrings}
\be
H=\frac{R_{11}}{2n}\left(
\left( 2\pi R_B w T_F \right)^2 + \left(\frac{w_{12}}{R_B}\right)^2
+p_i^2 +4\pi T_F\hat{N}\right) \ ,
\ee
where we have identified the decompactifying IIB radius
\be
R_B=\frac{l_P^3}{R_1 R_2}\gg l_P\ .
\ee

As soon as we put a single unit of R-R charge in the $U(1)$, i.e. the electric
field projects onto the long side in Figure~\ref{fig1}, the situation
changes in two ways; first, $H^{U(1)} \sim \mathcal{O}(T_D)$; second, the
string-like light excitations are now along a parallel direction to the electric
field  and are of $\mathcal{O}(T_F)$.
Considering
the tower of winding modes in the $U(1)$, we are not justified to ignore
the heavy excitations along $\tilde{\sigma}^2$ from the spectrum. The
heavy excitations present the spectrum of the $(q_1,q_2)$ string\cite{SL2Z}
\be
H=\frac{R_{11}}{2n}\left( p_i^2+M_{(q_1,q_2)}^2\right) \ ,
\ee
where $i=3,\ldots ,9$, and
\be
M_{(q_1,q_2)}^2= \left(2 \pi R_B w T_{(q_1,q_2)}\right)^2 +
\left( \frac{w_{12}}{R_B}\right)^2 + 4\pi T_{(q_1,q_2)} \hat{N} \ .
\ee
Off each level in this tower, we have a light spectrum
\be
H^{SU(n)}_{(1)} \rightarrow \frac{R_{11}}{2n} 4\pi \frac{T_F}{q_2} \hat{N}\ .
\ee

In general, it may be argued that a complete description of the system 
requires the full dynamics of the membrane theory with correspondingly more
complicated boundary conditions on the torus. Nevertheless, we claim that
this picture is enough to allow us to summarize the conclusions of the
analysis as follows.
Given a $U(1)$ electric field on the torus, we obtain the spectrum
of the $(q_1,q_2)$ bound strings by exciting the direction perpendicular
to the field in the $SU(n)$; configurations carrying
zero R-R charge have a tower of heavy excitations parallel to the electric
field pushed out of the spectrum
to $\mathcal{O}(T_F)$, leaving the spectrum of the fundamental IIB
string. Configurations carrying R-R
charge, through the non-trivial mechanism outlined above, have the bound
string spectrum
additionally swamped by light excitations parallel to the electric field.
We propose that these light excitations correspond 
to massive excitations of fundamental strings attached
to bound state of strings; even a single unit of R-R charge in the 
$(0,1)$ D-string generates this light spectrum; this is the signature of
the Dirichlet aspect of the underlying $U(1)$ vacuum. The fact that these
light excitations are of different 
$\sigma$ dependence than the ones in the $U(1)$
sector indicates further 
the different role played by these light strings. 

\section{Conclusions}

We have demonstrated, using energy arguments, the mechanism through which 
excitations in Matrix Theory, essentially excitations on a membrane, 
distribute themselves into the spectrum of bound IIB strings and the spectrum
of fundamental strings attached to D strings. We have also translated the 
connection between R-R charge and the Dirichlet aspect of the D-strings into
the language of the matrix YM, that of a non-trivial connection between
the $U(1)$ and the $SU(n)$ sectors. This feature arose from the mechanism
of winding a $U(1)$ scalar onto an oblique cycle, given
the charge content of the configuration; consequently, the periods of
the torus seen by the $SU(n)$ excitations contained the information about
the winding $U(1)$ scalar and the charges it carries. Finally, we note
that, whereas in~\cite{Waves}
the BPS condition picked out the string-like excitations on a cycle of the
membrane to be identified with the short spectrum of the bound
strings, in the Matrix Theory the energetics of the excitations naturally splits
the dynamics on the membrane 
into two sets of string-like spectra, one of which is mapped
onto the full non-BPS spectrum of the bound strings.

To further substantiate this proposal, a picture tracing 
the geometry and some of the 
dynamics of the massive excitations of the strings attached
to the state of bound strings 
is desirable; this may possibly explain the exact form
of the light excitations we have found.

\section{Acknowledgments}

I would like to thank E. Martinec for suggesting the problem and for
particularly illuminating discussions at various stages of the analysis. 
I would also like to acknowledge M. Li for useful suggestions.

\end{document}